# DISCRETE THERMODYNAMICS OF 2-LEVEL LASER – WHY *NOT* AND WHEN *YES*

B. Zilbergleyt,
System Dynamics Research Foundation, Chicago, USA
sdrf@ameritech.net

## INTRODUCTION

From a thermodynamic point of view, a laser transforms input energy flow into the output, and should obey the thermodynamic laws. As strange as it is, the only found approach was Einstein's thermodynamic prohibition on the ability of a 2-level laser to function [1], based on Boltzmann statistics and on the equal pump up – drop down probabilities for the excitable atoms. Einstein considered a laser as an isolated system, i.e. in absence of pumping energy; needless to say that a real lasing device is an open system. Nevertheless, if materialized in a concrete device, the 2-level laser should feature higher efficiency due to a decreased quantum defect and minimal energy dissipation as well as a wider choice of working media. It is still a very desirable yet not available thing.

Our objective was to outline a discrete thermodynamic model of a laser as an energy transforming open system. Discreteness in this case simply means usage of finite differences instead of derivatives, which leads to important advantages in complex systems treatment [2].

## THERMODYNAMIC CONSIDERATIONS

Recent research in the discrete thermodynamics of chemical systems [3] has brought out the conditions of equilibrium in an open system as a logistic map

(1) $$\ln[\Pi_j(\eta_j,0)/\Pi_j(\eta_j,\delta_j)] - \tau_j(1-\delta_j^p) = 0.$$

Its regular graphical solutions are conventional pitchfork bifurcation diagrams; the solution also may be plotted as a Dynamic Inverse Bifurcation Diagrams (DIBD, [3]) in coordinates "shift from equilibrium ($\delta$) vs. the shifting force ($F$)", with pitchfork limits 1 and 0 (Fig.1). A full set of them constitutes the system domain of states, and in this paper we will use exclusively DIBD.

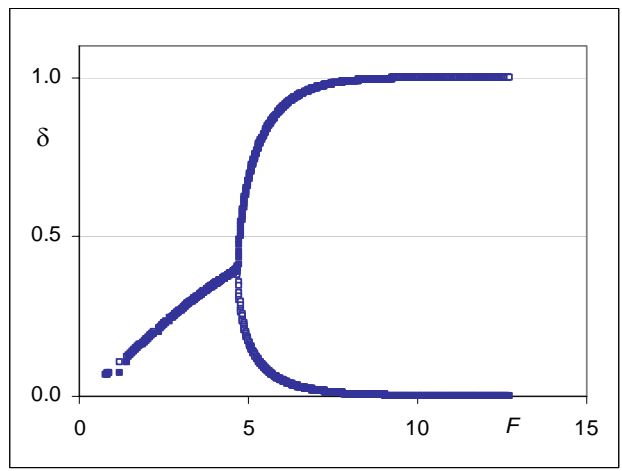

Fig.1. DIBD, shift from thermodynamic equilibrium vs. external thermodynamic force.
Reaction A+B=C, $\Delta g = -0.42$, $\eta = 0.37$ mol, $p=1$.

Map (1) corresponds to a process in the chemical system at p,T=const. The first, logarithmic term is merely reduced by RT change of Gibbs' free energy for the j-system, i.e. $\Delta g_j = \Delta g_j^0 + \ln[\Pi_j(\eta_j, \delta_j)]$, $\Delta g^0_j = -\ln K_j = -\ln[\Pi_j(\eta_j, 0)]$. The $\Delta g_j$ value is expressed via variable $\delta_j$ and parameter $\eta_j$ that are closely tied to *true thermodynamic equilibrium* (TdE), a state of an isolated system. As the system shifts from TdE, $\delta_j$ is a change of reaction coordinate, caused by thermodynamic force (TdF). A linear relationship between the TdF value and the shift power series $(\delta_j^0 + \delta_j^1 + \ldots + \delta_j^p)$ was assumed as one of the theory premises, p is a loosely defined indicator of the system complexity. The thermodynamic equivalent of transformation $\eta_j$ is an invariant, taking on the same value for all reaction participants, and is equal to the transformed (consumed or arrived *ab initio* to TdE) amounts of the reaction participants reduced by the corresponding stoichiometric units. This value memorizes the reaction change of Gibbs' free energy including its standard change as well as the system's initial composition and describes the system "strength" more convincingly than $\Delta g^0_j$. The second term of map (1) represents external thermodynamic force (TdF) with $\tau_j$ as a growth parameter of $\delta_j$. In the systems relevant to map (1) with p=1 the TdF numerically equals to $\tau_j$. For detailed explanations see [3,4].

Map (1) describes chemical (or quasi-chemical) equilibrium states. In isolated equilibrium TdF=0, then $\delta_j=0$ by definition, the second term turns to zero, and map (1) defines conditions of traditional equilibrium, coinciding with "true" thermodynamic equilibrium in isolated systems. Otherwise map (1) describes an *open equilibrium* (OpE), where the chemical equilibrium conditions do not match the same for "true" TdE. General graphical solutions to the map (1), with bifurcation diagrams, clearly show the triggering nature of such systems: a fast ascent to bifurcation point, a sharp rush of new branches up to unity and down to zero, and, typical for the triggering systems, bistability with possible oscillations between pitchfork branches.

Now we will show how to apply map (1) to a 2-level laser. Laser functioning is based on the population inversion between two levels with potential extreme values of the system shift from equilibrium: $\delta=0$ for the ground level that corresponds to non-excited, or equilibrium state, and $\delta=1$ for the excited, or laser level. This resemblance clearly brings laser to the class of triggering systems, and has prompted the present work. From the chemical (more exactly, quasi-chemical) point of view the laser represents an open system with a cyclic reaction – its ascending part is driven by external pumping power $\varepsilon_{in}$ and results in shifting a part of the laser population to the upper level:

(2) 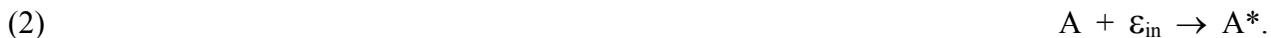 $A + \varepsilon_{in} \to A^*$.

Its descending part

(3) 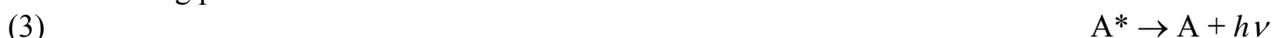 $A^* \to A + h\nu$

returns the population back to the ground level and takes responsibility for the light emission. The above-described parameters of map (1) take on a clear meaning: the thermodynamic equivalent of transformation $\eta$ represents the ground level equilibrium population; obviously, the shift $\delta$ is a portion of the laser population on the laser level at reaction (3) open equilibrium. A system complexity parameter may be roughly identified with the number of possible transition ways between the levels (p); in the suggested 2-level laser p=1. Taking into account these similarities, at a constant amount of the laser excitable dwellers (atoms, molecules, etc.), say, $n_A + n_{A^*} = 1$ mole, at p,T=const with stoichiometric coefficients (−1, 1) the Boltzmann-Gibbs distribution is

(4) $(1-\eta)/\eta = \exp(-h\nu/kT)$.

Multiplying both levels of the exponential power in (4) by the Avogadro number $N_A$, we obtain reduced by RT the standard change of Gibbs' free energy for reaction (3)

(5) $$\Delta g^0 = \Delta G^0/RT = -h\nu N_A/RT,$$

or, combining the universal constants into $\lambda_\nu = hN_A/R$

(6) $$\Delta g^0 = -\lambda_\nu \nu/T,$$

$\lambda_\nu = 47.99$ if the frequency is expressed in terahertz. Table I contains $\Delta g^0$ values for various colors.

Table I.
Standard changes of reduced Gibbs' free energy for reaction (3) at T=300K, visible light region.

| Color | Frequency interval or mean value | $\Delta g^0$, mean |
|---|---|---|
| shortwave uv | 1,542 THz | -246.720 |
| uv | 860 THHz | -137.440 |
| violet | ~ 700 to 790 THz | -119.200 |
| blue | ~ 600 to 700 THz | -104.000 |
| cyan | ~ 580 to 600 THz | -94.400 |
| green | ~ 530 to 580 THz | -88.800 |
| yellow | ~ 510 to 530 THz | -83.200 |
| orange | ~ 480 to 510 THz | -79.200 |
| red | ~ 405 to 480 THz | -70.400 |
| infra red | 394 THz | -63.040 |

The value of $\eta$ is easy to calculate from:

(7) $$\eta = 1/[1+\exp(\Delta g^0)].$$

It is clear that $\Delta g^0$ is always negative (the emitted energy leaves the system!); a glance at its values in Table I is enough to see that $\eta$ is very close to unity in absence of the external force: in isolated system Boltzmann-Gibbs statistics keeps the vast majority of the laser dwellers dormant on the ground level.

The reaction coordinate of reaction (3) at open equilibrium is shifted by $\delta$, and the populations become $\eta(1-\delta)$ for the ground level (reaction "product") and $[1-\eta(1-\delta)]$ for the upper level (reaction "reactant"); together with equation (6), that transforms map (1) into

(8) $$-\lambda_\nu \nu/T + \ln\{\eta(1-\delta)/[1-\eta(1-\delta)]\} + \tau(1-\delta) = 0.$$

The principal difference between this map and its parent classical expression (4) lays in explicit involvement of the pumping power that causes a non-zero positive reaction shift in (8). At $\delta=0$ map (8) turns to its TdE expression (4). Obviously $\delta=1$ is not achievable – map (8) cannot be balanced due to zero numerator under the logarithm, and the system is approaching this value asymptotically like in Fig.1. When $\delta$ is close to 1, solutions to map (8) experience deformation. The same should be mentioned regarding the close to 1 values of $\eta$, typical for the whole investigated frequency range (well recognized obstacle in equilibria calculations [5]). To avoid divergences at simulation, we kept $\eta$ restricted to 0.999 although the graphs outlook depends upon it (compare the rightmost part of Fig.2 with Fig.3).

**SIMULATION RESULTS**

We simulated open equilibria of the 2-level lasers for several light frequencies, including all of the colors in Table I; simulation results were graphically interpreted using DIBD, all graphs in this paper are plotted in shift vs. pumping force coordinates, both coordinate values are dimensionless. The pumping power takes on energy units being multiplied by RT (in our case RT=2.4404 kJ/mol).

Fig.2 shows DIBD for three values of $\eta$. On the dynamic diagrams new branches split almost at the $180^0$ angle at bifurcation point, promptly rushing to the trigger $\delta$ limits – up to unity and down to zero. That means that population inversion (approximately defined in the laser theory as 1:1 laser to ground populations ratio [2]) can be achieved only beyond bifurcation point. That prompts us to consider bifurcation point as a ***thermodynamic threshold of inversion***. Its coordinates may be easily found by simulation.

A close look at the results reveals new interesting features, to the best of the author's knowledge previously unknown. **First**, at lower e.m. frequencies, i.e. relatively low energies and low $\eta$ (left picture in Fig.2), the diagrams look like those reported in previous publications ([3], Fig.1). For larger $\eta$ the diagram outlook changes drastically: the vertical lines between the branches occur (middle in Fig.2), and for very large $\eta$ (right in Fig.2) the ascending part of the graph, the bifurcation point, and the lower branch all are tangibly shifted down. Beginning at a certain value of $\eta$, the diagram space between the branches is filled out with the line spectrum of alternating pumping up and spontaneous emission down impulses, in some parts of the spectrum following each other very rapidly (painted in blue). According to our data, the "no-line – line" spectrum border value of $\eta$ falls into 0.825–0.832 interval, corresponding approximately to 9… 10 THz.

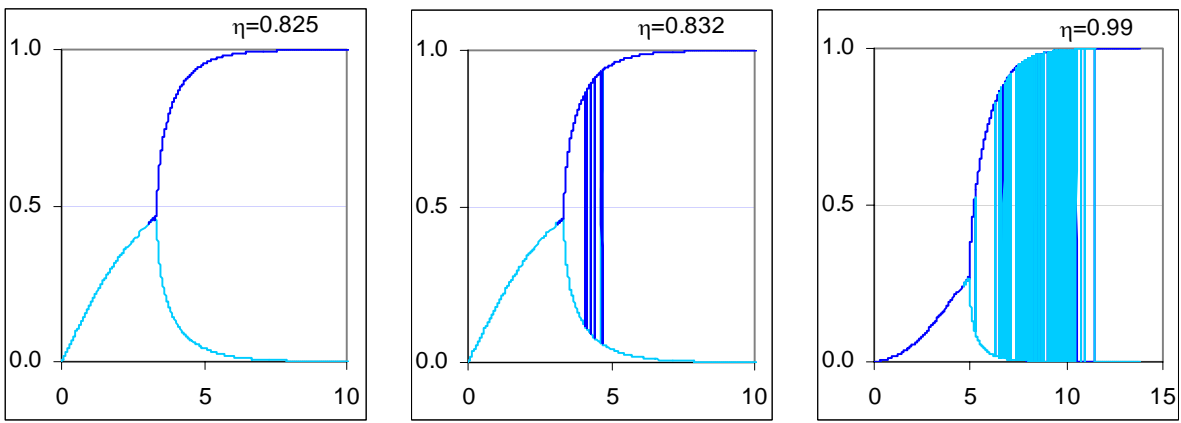

Fig.2. Dynamic Inverse Bifurcation Diagrams for reaction (3) at different $\eta$ values.

Fig.3 shows short wave UV dynamic diagram stretched horizontally. All of the up - down transitions frequency spectra (do not confuse with spectra of the emitted light!) are linear within the studied frequency range. In the laser terminology that means multiple, following each other transitions – the system is in open dynamic equilibrium, each move down is immediately followed by an opposite move of the same or similar shape and value.

**Second**, diagrams for various light waves – from infrared to short wave UV – are strikingly similar; this resemblance is caused by the large values of real $\eta$, all are not less than 0.999.

**Third**, though the simulated bifurcation point coordinates depend on the η value, it was possible to estimate their limit values for η→1 by the linear extrapolation:
− *thermodynamic threshold of inversion*     $P_{bp}$ =    6.59 ( $E_b$=16.08 kJ/mol at T=300K);
− *relative size of the laser level population*   $\delta_{bp}$ =    0.30 (population ratio 0.43);
− *system change of Gibbs' free energy*     $\Delta G_{bp}$ = −17.59 kJ/mol.

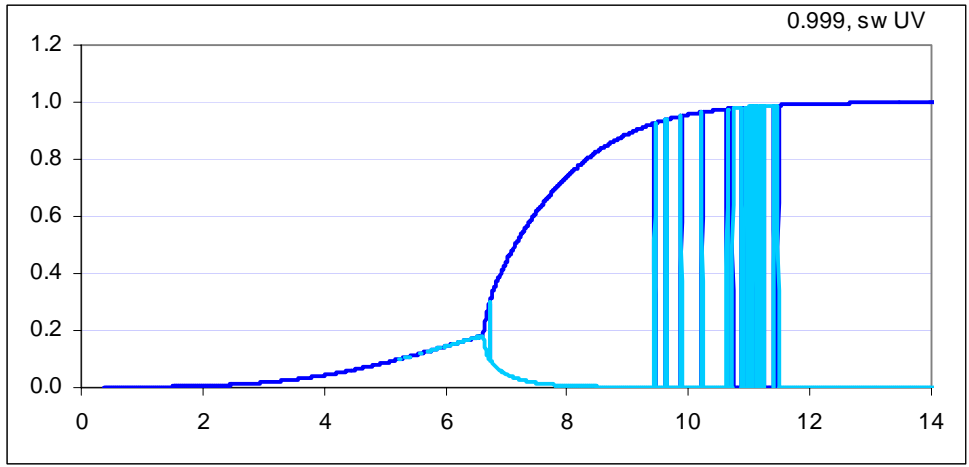

Fig.3. DIBD for the short wave UV laser, accepted η=0.999.

## 2-LEVEL LASER – WHY *NOT* AND WHEN *YES*

The above described linearity of the transition spectra, meaning close following of the alternating and overlapping opposite moves represent an ***independent, and exclusively based on discrete thermodynamics, additional proof of the Einstein's prohibition on the 2-level laser***. Pumped to the upper level laser dwellers cannot accumulate more energy from the pumping flow: a stress, caused by the pumping "pressure", forces the system to discharge itself by spontaneous or assisted irradiation of light with immediate re-pumping. Opposite transitions suppress each other, deterring powerful coherent monochromatic emission. The first of the title questions is answered thermodynamically!

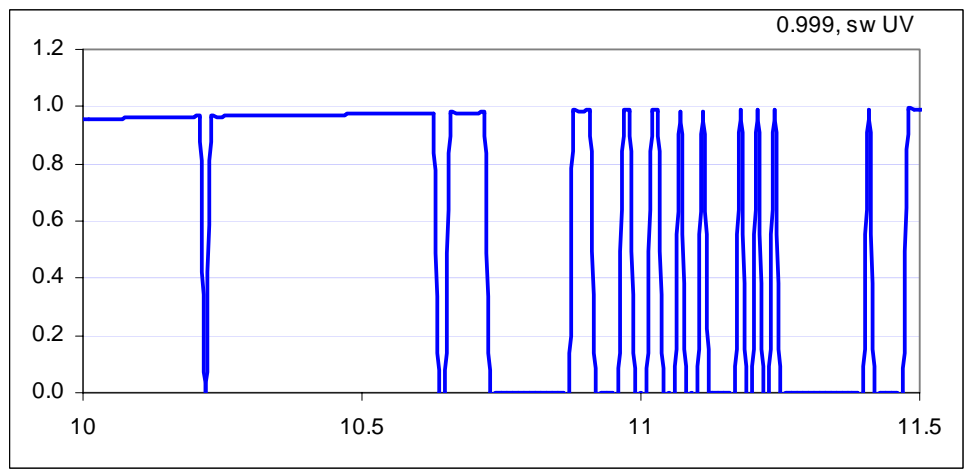

Fig.4. Enlarged fragment of the 2-level laser DIBD, short wave UV, shows changes of the laser level population as pumping force increases.

At the same time, another feature of the frequency transition spectra was discovered – a noticeable repetition of distinctive gaps in the emission frequency that shield the light emission between certain values of the pumping force. They are visible in Fig.2 and Fig.3; Fig.4 presents an enlarged fragment of the diagram showing the transitions from laser level down.

Could the gaps lead to a real 2-level laser? A combination of wide enough gap and a slight phase shift between opposite impulses along the pumping force axis may help to gain control over the emission by modulating the pumping force. That cannot be reached in series of closely following opposite narrow impulses. Indeed, we have detected some gaps as wide as 1.2 kJ/mol on the background of 25-35 kJ/mol level of the pumping power. They seem to be detectable experimentally, too.

**ON THE PRESENTATION OF LASER ACTION WITH BIFURCATION DIAGRAMS**

The idea that the laser-pumping curve is related to a pitchfork bifurcation diagram isn't new (e.g. [6], see Fig.5). A series of rather disordered bifurcation diagrams were presented in many other papers as well, e.g. in [7]. Most of them have discussed on a qualitative level the relations between the diagrams and operating features of lasers.

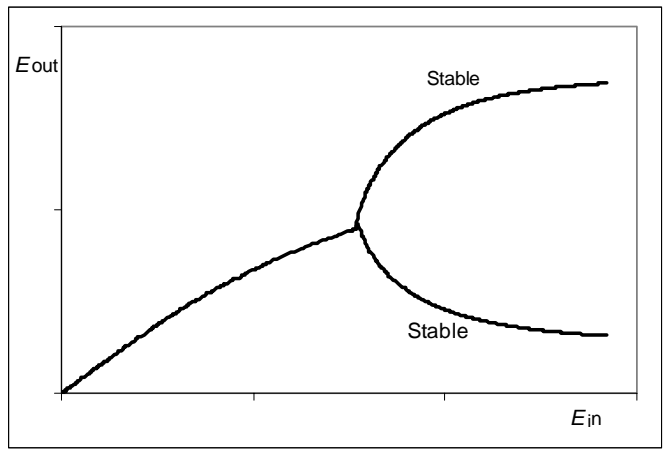

Fig.5. Output power vs. pump rate in bi-stable distributed feedback laser diode
(a mimicry to a drawing from [6]).

The difference between Fig.5 and Fig.3 is quite clear; results of our paper are quantitative and consistent in describing the full span switches between the laser and the ground levels. Whatever laser features follow from this model, they are of the pure thermodynamic origin, including what was known earlier on as well as what is new. To the best of our knowledge, such an approach has never been investigated before.

**SUMMARY**

Our goal was to outline a discrete thermodynamic model of the 2-level laser. Several ready to be explored right away features like an apparent fractal structure of the spectra, the gaps distribution on the pumping power axis, some others left untouched.

In real lasers, the retarding third, or the lasing level as well as the fourth one, close to the ground, help to bypass the Einstein's prohibition by separation of the opposite impulses in time instead of the above mentioned phase shift along the pumping axis. Developed in this work basic model may be extended to cover the most of conceivable lasers, and the time factor may be not the only tool to split the opposite impulses.

The system of discrete thermodynamics is organized in such a way that a generally expressed external thermodynamic impact can account for any external influence on reaction (3), including, e.g., any harvesting that decreases photon or A*− populations, like non-collective irradiation stimulated by the incoming photon adsorption and some others. Most of those effects can be rigorously formulated and incorporated into this model as thermodynamic factors.

**ACKNOWLEDGEMENTS**